\documentclass[fleqn,usenatbib]{mnras}

\usepackage[T1]{fontenc}

\DeclareRobustCommand{\VAN}[3]{#2}
\let\VANthebibliography\thebibliography
\def\thebibliography{\DeclareRobustCommand{\VAN}[3]{##3}\VANthebibliography}

\usepackage{graphicx}	
\usepackage{amsmath}	
\usepackage{amssymb}
\usepackage{xcolor}

\title[Formation of a magnetized hybrid star]{Formation of a magnetized hybrid star with a purely toroidal field from phase-transition-induced collapse}

\author[A. K. L. Yip et al.]{
Anson Ka Long Yip,$^{1}$\thanks{E-mail: klyip@phy.cuhk.edu.hk}
Patrick Chi-Kit Cheong$^{2,3,4}$
and Tjonnie Guang Feng Li$^{5,6}$
\\
$^{1}$Department of Physics, The Chinese University of Hong Kong, Shatin, N.T., Hong Kong\\
$^{2}$Department of Physics \& Astronomy, University of New Hampshire, 9 Library Way, Durham NH 03824, USA\\
$^{3}$Center for Nonlinear Studies, Los Alamos National Laboratory, Los Alamos, NM 87545, USA\\
$^{4}$Department of Physics, University of California, Berkeley, Berkeley, CA 94720, USA\\
$^{5}$Institute for Theoretical Physics, KU Leuven, Celestijnenlaan 200D, B-3001 Leuven, Belgium\\
$^{6}$Department of Electrical Engineering (ESAT), KU Leuven, Kasteelpark Arenberg 10, B-3001 Leuven, Belgium
}

\date{Accepted XXX. Received YYY; in original form ZZZ}

\pubyear{2024}

\begin{document}
\label{firstpage}
\pagerange{\pageref{firstpage}--\pageref{lastpage}}
\maketitle

\begin{abstract}
Strongly magnetized neutron stars are popular candidates for producing detectable electromagnetic and gravitational-wave signals.
Gravitational collapses of neutron stars triggered by a phase transition from hadrons to deconfined quarks in the cores could also release a considerable amount of energy in the form of gravitational waves and neutrinos.
Hence, the formation of a magnetized hybrid star from such a \emph{phase-transition-induced collapse} is an interesting scenario for detecting all these signals.
These detections may provide essential probes for the magnetic field and composition of such stars.
Thus far, a dynamical study of the formation of a magnetized hybrid star from a \emph{phase-transition-induced collapse} has yet to be realized.
Here, we investigate the formation of a magnetized hybrid star with a purely toroidal field and its properties through dynamical simulations.
We find that the maximum values of rest-mass density and magnetic field strength increase slightly and these two quantities are coupled in phase during the formation.
We then demonstrate that all microscopic and macroscopic quantities of the resulting hybrid star vary drastically when the maximum magnetic field strength goes beyond a threshold of $\sim 5 \times 10^{17}$ G but they are insensitive to the magnetic field below this threshold. 
Specifically, the magnetic deformation makes the rest-mass density drop significantly, suppressing the matter fraction in the mixed phase.
These behaviors agree with those in the equilibrium models of previous studies.
Therefore, this work provides a solid support for the magnetic effects on a hybrid star.
\end{abstract}

\begin{keywords}
(magnetohydrodynamics) MHD --
stars: neutron --
stars: magnetic field -- 
methods: numerical -- 
software: simulations
\end{keywords}



\section{Introduction} \label{sec:intro}

Neutron stars are natural laboratories for studying physics under extreme conditions, which terrestrial experiments cannot reproduce. 
On the one hand, the density of a neutron star reaches above the nuclear saturation density $\rho_{0} \sim 2.8 \times 10^{14}$ g cm$^{-3}$, at which the canonical atomic structure of matter is disrupted. 
The detailed microphysics and the concerning equation of state at supra-nuclear densities still remain elusive. Exotic matter, such as deconfined quark matter and hyperons, could exist in this ultradense regime (See e.g. \citealt{2018ASSL..457.....R} for a review).

Several studies have long proposed compact stars that are partly or wholly composed of deconfined quark matter \citep{1970PThPh..44..291I,1971PhRvD...4.1601B,1984PhRvD..30..272W}. 
These stars are typically interpreted as the products of the phase transition of the hadrons in the original neutron stars. 
In particular, when the density inside a neutron star reaches a threshold, a phase transition converting hadrons into deconfined quarks could happen.
If this phase transition only occurs in the stellar core, the resulting star is usually called a `hybrid star'. 
Since the equation of state describing deconfined quark matter is generally ``softer'' than that of hadronic matter, a hybrid star generally has a smaller radius and higher compactness than the progenitor neutron star.
The softening of the equation of state due to the phase transition in the stellar core could trigger a gravitational collapse and release gravitational potential energy of the order of $\sim 10^{52}$ erg.
Significant portions of the released energy could give rise to the emission of neutrinos and gravitational waves. 
Detecting these signals provides evidence of deconfined quark matter. 
Newly born neutron stars in supernovae and accreting neutron stars in binary systems are possible hosts for such a \emph{phase-transition-induced collapse} (See e.g. \citealt{1999JPhG...25..195W,2009MNRAS.392...52A} for reviews).

The major difficulty in studying these \emph{phase-transition-induced collapses} is that the exact nature of the phase transition is still poorly known and actively investigated.
In the literature, two distinct scenarios can be found regarding the process of phase transition: a slow deflagration \citep{2007ApJ...659.1519D,2018IJMPE..2750083M} and a fast detonation \citep{2006PhRvC..74f5804B,2007PhRvC..76e2801B,2015PhRvC..91e5806M}.
In particular, \citet{2007ApJ...659.1519D} examined the possible modes of burning of hadronic matter into quark
matter using relativistic hydrodynamics and a microphysical equation of state.
They showed that the conversion process consistently takes the form of a deflagration and never a detonation.
They also estimated the increase in the phase transition propagation velocity due to hydrodynamical instabilities and demonstrated that these instabilities are inadequate to convert the deflagration into a detonation.
Consequently, the conversion process consistently exhibits deflagration and does not transform into a detonation.
On the other hand, \citet{2006PhRvC..74f5804B} treated the phase transition as a two-step process,
Initially, the conversion involves transforming hadronic matter into two-flavor u and d quark matter, followed by the subsequent transition to strange quark matter (u, d, and s quarks) in a second step.
They employed relativistic hydrodynamics to calculate the propagation velocity of the first front. They observed the development of a detonation wave in the hadronic matter during this initial stage. 
Once this front passes through and leaves behind two-flavor matter, a second front arises, facilitating the transformation of the two-flavor matter into three-flavor matter through weak interactions. 
The time scale for the second conversion is $\sim$ 100 s, while the first step takes about $\sim$ 1 ms.

For a detonation-like conversion, the conversion front travels at a sound speed of $\sim$ 0.3-0.5$c$, where $c$ is the speed of light.
If the radius of the stellar core undergoing the phase transition is $\sim$ 5 km, then the timescale for such conversion is $\sim$ 0.05 ms, much smaller than that of the dynamical timescale of the star.
In this context, \citet{2006ApJ...639..382L} took the first step to study the \emph{phase-transition-induced collapse} through Newtonian hydrodynamics simulations.
They mimicked a phase transition that occurs instantaneously in the stellar core by changing the equation of state in the initial time slice of the simulations.
Instead of detailed modeling of the short timescale conversion process, they focused on the dynamics of subsequent collapse due to the phase transition.
\citet{2009MNRAS.392...52A} extended this work by considering the effect of general relativity and introducing a finite timescale for the initial phase transformation.
They showed that introducing this finite timescale only causes small changes in the dynamics.

On the other hand, neutron stars have the strongest magnetic field found in the Universe. 
Dipole spin-down models allow for the estimation of the surface magnetic field strength of neutron stars. 
With the surface field strength $\mathcal{B}_\mathrm{s}$, we can classify neutron stars into millisecond pulsars with $\mathcal{B}_\mathrm{s} \sim 10^{8 - 9}$ G, classical pulsars with $\mathcal{B}_\mathrm{s} \sim 10^{11 - 13}$ G, and magnetars with $\mathcal{B}_\mathrm{s} \sim 10^{14 - 15}$ G.
Although there is still no direct observation of the interior magnetic field of neutron stars, arguments based on the virial theorem suggest that it could reach $10^{18}$ G (see e.g. \citealt{2010PhRvC..82f5802F}, \citealt{1991ApJ...383..745L}, \citealt{1989ApJ...342..958F}, and \citealt{2001ApJ...554..322C}). 
Furthermore, binary neutron star simulations have demonstrated that the local maximum magnetic field can be amplified up to $\sim 10^{17}$ G during the merger \citep{2006Sci...312..719P,2015PhRvD..92f4034K,2015PhRvD..92l4034K,2020PhRvD.102j3006A}.

Highly magnetized neutron stars are promising candidates for explaining some puzzling astronomical phenomena, including soft gamma-ray repeaters and anomalous X-ray pulsars \citep{1998Natur.393..235K,1999ApJ...510L.111H,1995ApJ...442L..17M,2000A&A...361..240M,1995A&A...299L..41V}.
Moreover, neutron stars can be deformed by the magnetic field, depending on the geometry of the magnetic field. 
A purely toroidal field induces prolateness \citep{2008PhRvD..78d4045K,2009ApJ...698..541K,2012MNRAS.427.3406F}, while a purely poloidal field causes oblateness to neutron stars \citep{1995A&A...301..757B,2001A&A...372..594K,2012PhRvD..85d4030Y}. 
These magnetic-field-induced distortions make rotating neutron stars possible sources for the emission of detectable continuous gravitational waves \citep{1996A&A...312..675B}.
However, the actual field geometry inside neutron stars is still unknown. 
Stability analyses of magnetized stars suggest that simple geometries are subjected to instabilities \citep{1957PPSB...70...31T,1973MNRAS.161..365T,1973MNRAS.163...77M,1974MNRAS.168..505M,1973MNRAS.162..339W}. 
Magnetohydrodynamics simulations propose a mixed configuration of toroidal and poloidal fields as the most favored configuration \citep{2006A&A...450.1077B,2006A&A...450.1097B,2009MNRAS.397..763B}. 
This configuration is usually referred to as the `twisted torus'.

Deconfined quarks and strong magnetic fields are expected to be present inside neutron stars, so studying magnetized hybrid stars is necessary to probe the combined effects of these two features.
Previous studies have investigated the properties of magnetized hybrid stars by constructing equilibrium models (e.g. \citealt{2009JPhG...36k5204R,2012EPJA...48..189D,2015JPhCS.607a2013I,2015MNRAS.447.3785C,2016MNRAS.463..571F,2016MNRAS.456.2937F,2022MNRAS.512..517M,2023ApJ...943...52R}). 
In particular, \citet{2015MNRAS.447.3785C} and \citet{2016MNRAS.456.2937F} have demonstrated that the pure field contribution to the energy–momentum tensor primarily contributes to the macroscopic properties of magnetized hybrid stars. 
In contrast, the magnetic effects in the equation of state and the field-matter interactions have negligible effects on these properties. 
Moreover, a magnetic field reduces the central density and prevents the appearance of quark matter. 
However, a dynamical study of a magnetized hybrid star has yet to be realized.
Besides, the effects on the configurations of a magnetized star (e.g. profiles of rest-mass density and magnetic field) due to a \emph{phase-transition-induced collapse} is still unknown.
Therefore, it is indispensable to confirm whether the magnetic effects found in previous studies can still accurately describe the magnetized hybrid stars that are dynamically formed by \emph{phase-transition-induced collapses}. 

In this work, based on the framework introduced by \cite{2006ApJ...639..382L}, we numerically study the formation of a magnetized hybrid star from \emph{phase-transition-induced collapse} through general relativistic magnetohydrodynamics simulations.
We have discussed the preliminary results of this work in \citet{Yip:2023qkh}.
Specifically, we first construct initial magnetized neutron star models before the phase transition by the open-sourced code \texttt{XNS} \citep{2011A&A...528A.101B,2014MNRAS.439.3541P,2015MNRAS.447.2821P,2017MNRAS.470.2469P,2020A&A...640A..44S}.
We then initiate the gravitational collapse of a magnetized neutron star by changing the equation of state in the central region to a ``softer'' equation of state, which considers deconfined quarks.
This \emph{phase-transition-induced collapse} gives a magnetized hybrid star as the final product.
The whole process is simulated using the new general relativistic magnetohydrodynamics code \texttt{Gmunu} \citep{2020CQGra..37n5015C,2021MNRAS.508.2279C,2022ApJS..261...22C}.
The details of the initial neutron star models, hybrid star models and evolutions are described in Section \ref{sec:num_method}.
The results of the formation process and the properties of the resulting star are presented in Sections \ref{sec:dynamics} and \ref{sec:properties} respectively. 
Finally, we provide the conclusions in Section \ref{sec:conclusions}.

\section{Numerical methods}\label{sec:num_method}
\subsection{Initial neutron star models}
We construct the non-rotating magnetized neutron star with a purely toroidal field equilibrium models in axisymmetry by the open-sourced code \texttt{XNS}  \citep{2011A&A...528A.101B,2014MNRAS.439.3541P,2015MNRAS.447.2821P,2017MNRAS.470.2469P,2020A&A...640A..44S}. 
These equilibrium models serve as initial data for our simulations.

We adopt the same way of parametrization as \citet{2014MNRAS.439.3541P} for the equation of state and the toroidal magnetic field due to two reasons.
Firstly, since the magnetic effects on equilibrium models are well-studied by \citet{2014MNRAS.439.3541P}, their models are good reference points to compare with our magnetized hybrid star models formed after \emph{phase-transition-induced collapses}.
Secondly, the main focus of this study is the \emph{phase-transition-induced collapse} triggered by the softening of the equation of state instead of the effects coming from the details of an equation of state.
Hence, we adopt a simplified equation of state for modeling the neutron stars. 

The initial neutron star models are constructed with a polytropic equation of state,
    \begin{equation}
    P=K \rho^\gamma,
    \end{equation}
where $P$ is the pressure, $\rho$ is the rest-mass density and we choose a polytropic constant $K=1.6 \times 10^5$ cm$^5$ g$^{-1}$ s$^{-2}$ (which equals to 110 in the unit of $c=G=M_{\odot}=1$) and a polytropic index $\gamma=2$. 

We specify the specific internal energy $\epsilon$ on the initial time-slice by
    \begin{equation}
    \epsilon=\frac{K}{\gamma-1} \rho^{\gamma-1}.
    \end{equation}

We adopt a magnetic polytropic law for the toroidal magnetic field
    \begin{equation}
        \mathcal{B}_{\phi}=\alpha^{-1}K_{\mathrm{m}}(\rho h\varpi^2)^m
    \end{equation}
where $\alpha$ is the laspe function, $K_{\mathrm{m}}$ is the toroidal magnetization constant, $h$ is the specific enthalpy, $\varpi^2=\alpha^2\psi^4r^2\sin^2\theta$, $\psi$ is the conformal factor, $(r,\theta)$ are the radial and angular coordinates in 2D spherical coordinates, and $m\geq1$ is the toroidal magnetization index.

The toroidal magnetic field used in our models is equivalent to a magnetic field pointing in the $\phi$-direction in spherical coordinates (see e.g. \citealt{2017MNRAS.466.1330H} for a visualization of a toroidal magnetic field). 
As mentioned in Section \ref{sec:intro}, this purely toroidal field configuration is expected to be unstable. 
However, since our simulations adopt 2D axisymmetry, the magnetic instabilities due to this configuration are suppressed.
Despite this, many models with mixed fields in the literature still exhibit instability (see e.g. \citealt{2023PhRvD.108h4006S}). 
Since the stable magnetic field configuration is still uncertain, this study serves as an initial exploration into the formation of a magnetized hybrid star with a purely toroidal field configuration.

In total, 9 models are constructed, where `REF' is the non-magnetized reference model and the remaining 8 neutron star models are magnetized.
They are part of the models used in \citet{2022CmPhy...5..334L}.
Because we do not intend to perform a comprehensive study of neutron stars with different masses in this work, all models have a fixed baryonic mass $M_0=1.68$ $M_\odot$. 
Latest observations of high-mass neutron stars are J0348+0432 at $M = 2.01 \pm 0.04 M_\odot$ \citep{2013Sci...340..448A}, PSR J0740+6620 at $M = 2.08 \pm 0.07 M_\odot$ \citep{2021ApJ...915L..12F}, and PSR J0952-0607 at $M = 2.35 \pm 0.17 M_\odot$ \citep{2022ApJ...934L..17R}.
These observations imply that the maximum mass of a neutron star should be above $2.0 M_\odot$. Hence, the adopted masses of our models are within the observational constraints on the maximum mass of a neutron star.
Also, the 8 magnetized models have the same toroidal magnetization index $m=1$ but different values of toroidal magnetization constant $K_{\mathrm{m}}$. 
They are arranged in the order of increasing maximum magnetic field strength $\mathcal{B}_\mathrm{max}$, where the model `T1K1' has the lowest strength, and `T1K2' has the second-lowest strength, so on and so forth.
(`T1' represents the toroidal magnetization index $m=1$ and `K' indicates the toroidal magnetization constant $K_{\mathrm{m}}$).
The configuration of these models allows a phase transition that occurs inside the stellar core.
Table~\ref{table1} summarizes the detailed properties of all 9 models.
It is unlikely to find the strongest magnetic field $\mathcal{B}_\mathrm{max} \sim 10^{17}$ G of our models on the \emph{surface} of ordinary pulsars and magnetars. 
Nonetheless, this ultra-high field strength could exist inside the stars because the toroidal magnetic fields considered are enclosed within the stars.
Besides, as mentioned in Section \ref{sec:intro}, arguments based on the virial theorem suggest that the interior magnetic field of a neutron star could reach $10^{18}$ G (see e.g. \citealt{2010PhRvC..82f5802F}, \citealt{1991ApJ...383..745L}, \citealt{1989ApJ...342..958F}, and \citealt{2001ApJ...554..322C}).

\begin{table}
	\centering
	\caption{\label{table1} Properties of the 9 initial neutron star models constructed by the \texttt{XNS} code.
	All numerical values are rounded off to two decimal places.
	$\rho_{\mathrm{c}}$ is the central rest-mass density, $M_{\mathrm{g}}$ is the gravitational mass, $r_{\mathrm{e}}$ is the equatorial radius, and $\mathcal{B}_{\mathrm{max}}$ is the maximum toroidal field strength inside the neutron star. 
    All the models have a fixed baryonic mass $M_{\mathrm{0}}=1.68$ $M_{\odot}$ and the 8 magnetized models also have the same toroidal magnetization index $m=1$.}
	\begin{tabular}{cccccccc}
		 Model & $\rho_{\mathrm{c}}$ & $M_{\mathrm{g}}$ & $r_\mathrm{e}$ & $\mathcal{B}_{\mathrm{max}}$\\
		 & ($10^{14}$ g cm$^{-3}$) & ($M_{\odot}$) & (km) & ($10^{17}$ G)\\
		\hline
		 REF & 8.56 & 1.55  & 11.85 & 0.00\\
		 T1K1 & 8.56 & 1.55 & 11.85 & $3.45\times10^{-2}$\\
		 T1K2 & 8.56 & 1.55 & 11.85 & $6.89\times10^{-2}$\\
		 T1K3 & 8.57 & 1.55 & 11.85 & $3.44\times10^{-1}$\\
		 T1K4 & 8.63 & 1.55 & 11.92 & 1.36\\
		 T1K5 & 8.81 & 1.56 & 12.15 & 2.63\\
		 T1K6 & 9.10 & 1.58 & 14.43 & 5.52\\
		 T1K7 & 8.81 & 1.59 & 16.21 & 6.01\\
		 T1K8 & 8.27 & 1.60 & 18.64 & 6.14\\
	\end{tabular}

\end{table}

\subsection{Hybrid star models and evolution}
As mentioned in Section \ref{sec:intro}, the main focus of this study is the \emph{phase-transition-induced collapse} triggered by the softening of the equation of state.
Hence, we adopt the simplified equation of state used in \citet{2009MNRAS.392...52A} for modeling the hybrid stars.
Since the effects coming from the details of an equation of state (e.g. finite temperature and magnetic effects considered in an equation of state) could also affect the internal dynamics of the star, we leave a complete study focusing on these effects for future work.
In particular, we assume that the phase transition occurs instantaneously in the initial time slice and is induced by changing the original polytropic equation to a ``softer'' equation of state for describing hybrid stars.
The assumption for the phase transition based on \citet{2006ApJ...639..382L} focuses on the dynamics after the phase-transition-induced collapse rather than the details during the short timescale conversion process.
In addition, \citet{2009MNRAS.392...52A} have demonstrated that introducing a finite timescale for the phase transition only causes small changes in the dynamics, so this treatment is suitable for describing a fast-detonation type of phase transition.

The MIT bag model equation of state introduced by \citet{johnson1975bag} has been widely used to model quark matter inside compact stars (see e.g. \citealt{weber1999quark,glendenning2012compact} for a review). 
The MIT bag model equation of state for massless and non-interacting quarks is given by
   \begin{equation}
    P_{\mathrm{q}}=\frac{1}{3}(e-4 B),
    \end{equation}
where $P_{\mathrm{q}}$ is the pressure of quark matter, $e$ is the total energy density and $B$ is the bag constant.

For the normal hadronic matter, we adopt an ideal gas type of equation of state for the evolution
    \begin{equation}
    P_{\mathrm{h}}=(\gamma-1) \rho \epsilon
    \end{equation}
where $P_{\mathrm{h}}$ is the pressure of hadronic matter and $\gamma$ is kept to be 2.

Either two or three parts constitute the hybrid star formed after the phase transition: (i) a hadronic matter region with a rest-mass density below the lower threshold density $\rho_\mathrm{hm}$, (ii) a mixed phase of the deconfined quark matter and hadronic matter for the region with a rest-mass density in between the lower threshold density $\rho_\mathrm{hm}$ and the upper threshold density $\rho_\mathrm{qm}$, and (iii) a region of pure quark matter phase with a rest-mass density beyond $\rho_\mathrm{qm}$ (based on the maximum density achieved, this may or may not be present in practice). 
With this picture, the equation of state for hybrid stars is given by
\begin{equation}
P= \begin{cases}P_{\mathrm{h}} & \text { for } \rho<\rho_{\mathrm{hm}}, \\ \alpha_\mathrm{q} P_{\mathrm{q}}+(1-\alpha_\mathrm{q}) P_{\mathrm{h}} & \text { for } \rho_{\mathrm{hm}} \leq \rho \leq \rho_{\mathrm{qm}}, \\ P_{\mathrm{q}} & \text { for } \rho_{\mathrm{qm}}<\rho,\end{cases}
\label{eqn6}
\end{equation}
where 
\begin{equation}
\alpha_\mathrm{q}=1-\left(\frac{\rho_{\mathrm{qm}}-\rho}{\rho_{\mathrm{qm}}-\rho_{\mathrm{hm}}}\right)^\delta
\end{equation}
is a scale factor to quantify the relative contribution due to hadronic and quark matters to the total pressure in the mixed phase. 
The exponent $\delta$ adjusts the pressure contribution due to quark matter.
We set 3 values of $\delta \in \{1, 2, 3\}$ to investigate the dynamical effects of varying quark matter contributions. 
We choose $\rho_\mathrm{hm}= 6.97 \times 10^{14}$ g cm$^{-3}$, $\rho_\mathrm{qm}= 24.3 \times 10^{14}$ g cm$^{-3}$ and $B^{1/4}=170$ MeV.
The above equation of state model for hybrid stars (including the parameters $\alpha_\mathrm{q}, \delta, \rho_\mathrm{hm}, \rho_\mathrm{qm}$) is similar to that in \citet{2009MNRAS.392...52A}.

We employ the new general relativistic magnetohydrodynamics code \texttt{Gmunu} \citep{2020CQGra..37n5015C,2021MNRAS.508.2279C,2022ApJS..261...22C} to evolve the stellar models in dynamical spacetime. 
\texttt{Gmunu} solves the Einstein equations in the conformally flat condition approximation based on the multigrid method.

We perform 2D ideal general-relativistic magnetohydrodynamics simulations in axisymmetry with respect to the $z$-axis and equatorial symmetry using cylindrical coordinates $(R,z)$. 
The computational domain covers [0,100] for both $R$ and $z$, with the base grid resolution $N_{R} \times N_{z} = 32 \times 32$ and allowing 6 AMR levels (effective resolution $= 1024 \times 1024$).
The refinement criteria of AMR is the same as that in \cite{2021MNRAS.508.2279C,2022CmPhy...5..334L}.
Our simulations adopt TVDLF approximate Riemann solver \citep{1996JCoPh.128...82T}, 3rd-order reconstruction method PPM \citep{1984JCoPh..54..174C} and 3rd-order accurate SSPRK3 time integrator \citep{1988JCoPh..77..439S}. 
The region outside the star is filled with an artificial low-density `atmosphere' with rest-mass density $\rho_\mathrm{atm} \sim 10^{-10} \rho_\mathrm{c}(t=0)$. 
Since we are restricted to purely toroidal field models and axisymmetry for the simulations, we do not use any divergence cleaning method.

\section{Formation dynamics}\label{sec:dynamics}
For each of the 9 equilibrium models, we perform simulations for three times, once for each value of the exponent $\delta \in \{1, 2, 3\}$. 
Consequently, $9 \times 3 = 27$ simulations are performed in total.

Since all simulations exhibit the same behavior, we take one of them as an example to describe the features of the formation dynamics.
Here, we choose the simulation with an initial maximum magnetic field strength $\mathcal{B}_\mathrm{max} = 5.52 \times 10^{17}$ G (i.e. Initial model T1K6) and an exponent $\delta=3$.
The exponent $\delta=3$ corresponds to a more substantial phase transition effect, which favors the demonstration of the formation dynamics.

Due to the appearance of the quark matter in the stellar core, the pressure in the core reduces, and the star starts to contract.
When the density in the core rises to a level where the pressure gradient balances the gravitational force, the star then bounces back and forth.
Since the ideal gas equation of state is employed for the evolution of the hadronic matter, shock waves form at every pulsation and dissipate the kinetic energy of the oscillation into thermal energy.
When these shock waves reach the stellar surface, a small amount of matter is ejected from the star and goes into the artificial low-density atmosphere until it crosses the outer numerical boundary (see e.g. \citealt{2009PhRvD..79b4017C,2011PhRvD..84d4012T}).
Eventually, the oscillation is damped due to this shock heating effect (also numerical dissipation discussed in Appendix \ref{sec:res_study}), and the star reaches a new equilibrium configuration of a hybrid star.

\begin{figure}
    \centering
    \includegraphics[width=\columnwidth, angle=0]{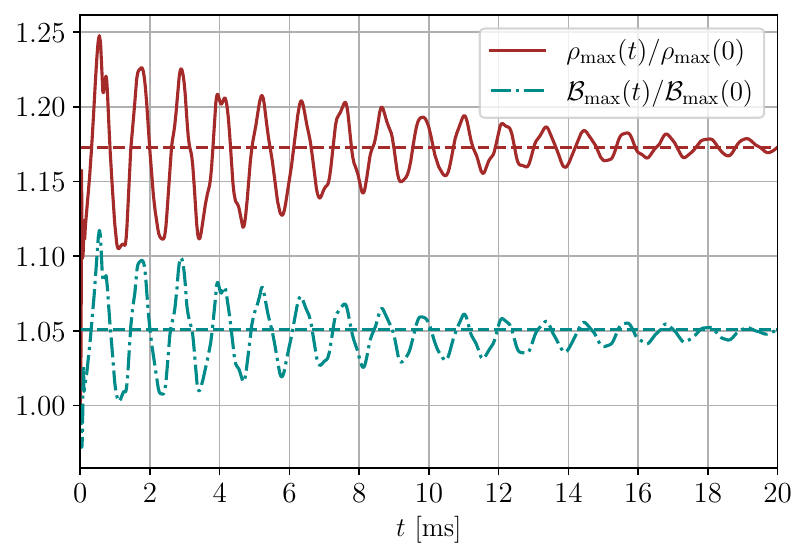}
    \caption{
            The time evolution of the maximum values of the rest-mass density $\rho_\mathrm{max}(t)$ (brown solid line) and the magnetic field strength $\mathcal{B}_\mathrm{max}(t)$ (dark cyan dash-dotted line) relative to their initial values $\rho_\mathrm{max}(0)$ and $\mathcal{B}_\mathrm{max}(0)$ for the simulation with the initial model T1K6 and the exponent $\delta=3$.
            Model T1K6 has an initial maximum magnetic field strength $\mathcal{B}_\mathrm{max} = 5.52 \times 10^{17}$ G and $\delta$ is an exponent describing the pressure contribution due to quark matter in the mixed phase.
            Dashed lines are the equilibrium values of the two quantities obtained at $t=20$ ms. 
            Similar damped oscillatory behaviors are observed for both quantities and the star is relaxed into a new equilibrium configuration after the \emph{phase-transition-induced collapse}.
            Importantly, these two quantities are coupled in phase during the formation process. 
            Moreover, after reaching the peak values at $t \sim 0.5$ ms, the oscillation amplitudes of $\rho_\mathrm{max}(t)/\rho_\mathrm{max}(0)$ and $\mathcal{B}_\mathrm{max}(t)/\mathcal{B}_\mathrm{max}(0)$ are reduced by a factor of $e^{-1}$ at $t\sim 8.5$ ms and $t\sim 6$ ms respectively.} 
    \label{fig1}	
\end{figure}

\begin{figure}
    \centering
    \includegraphics[width=\columnwidth, angle=0]{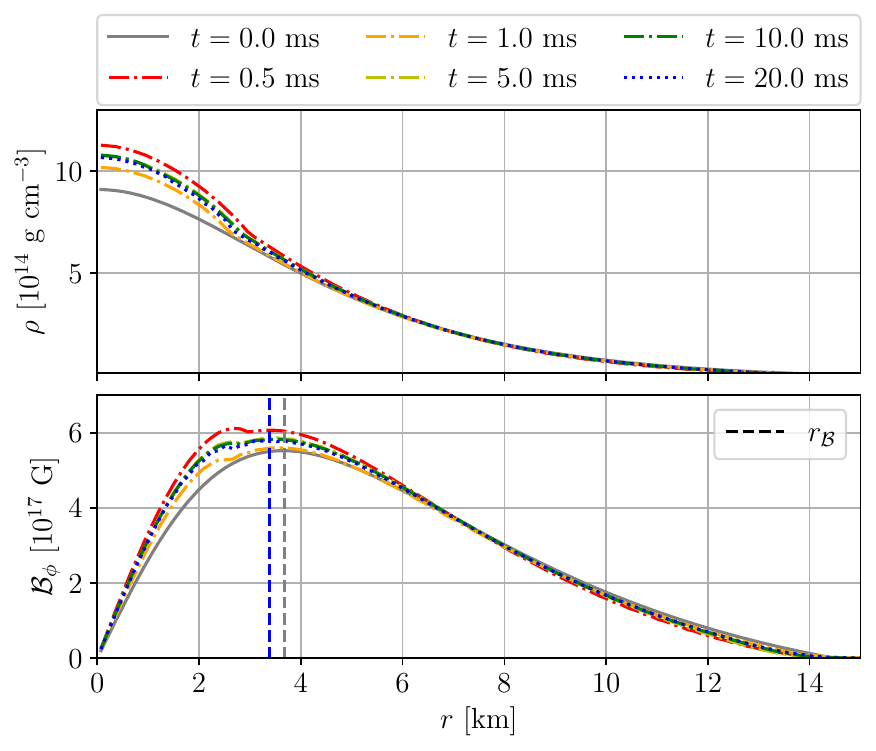}
    \caption{
	        The radial profile of the rest-mass density $\rho(r)$ (top panel) and the toroidal magnetic field strength $\mathcal{B}_{\phi}(r)$ (bottom panel) in the equatorial plane (i.e. polar angle $\theta= \pi/2$) for the simulation with the initial model T1K6 and the exponent $\delta=3$ at $t=0$ ms (gray solid lines), 0.5 ms (red dash-dotted lines), 1.0 ms (orange dash-dotted lines), 5.0 ms (yellow dash-dotted lines), 10.0 ms (green dash-dotted lines), and 20.0 ms (blue dotted lines).
            Model T1K6 has an initial maximum magnetic field strength $\mathcal{B}_\mathrm{max} = 5.52 \times 10^{17}$ G and $\delta$ is an exponent describing the pressure contribution due to quark matter in the mixed phase.
            The dashed lines in the lower panel represent the maximum magnetic field strength position $r_{\mathcal{B}}$.
            The \emph{phase-transition-induced collapse} causes the star initially oscillates around a new equilibrium configuration for both $\rho(r)$ and $\mathcal{B}_{\phi}(r)$ and the resulting hybrid star eventually obtains a slightly higher central rest-mass density and maximum magnetic field strength.
            Also, the magnetic field inside the star becomes more concentrated towards the core with a shift of $r_{\mathcal{B}}$ to smaller values. 
            Moreover, these new configurations of $\rho(r)$ and $\mathcal{B}_{\phi}(r)$ are obtained at $t=10$ ms and remain the same until at least $t=20$ ms.
	         }
    \label{fig2}	
\end{figure}

This kind of damped oscillatory behavior can be clearly seen in Fig.~\ref{fig1}, which shows the time evolution of the maximum values of the rest-mass density $\rho_\mathrm{max}(t)$ (brown solid line) and the magnetic field strength $\mathcal{B}_\mathrm{max}(t)$ (dark cyan dash-dotted line) relative to their initial values $\rho_\mathrm{max}(0)$ and $\mathcal{B}_\mathrm{max}(0)$.
$\rho_\mathrm{max}$ is always located at the centre of the star, but $\mathcal{B}_\mathrm{max}$ is at a position away from the centre (See Fig. \ref{fig2} for details).
The equilibrium values obtained at $t=$ 20 ms are plotted with dashed lines. 
We observe similar damped oscillatory behaviors for both quantities and and the star is relaxed into a new equilibrium configuration after the \emph{phase-transition-induced collapse}.
Importantly, these two quantities are coupled in phase during the formation process.
Moreover, after reaching their peak values at $t \sim 0.5$ ms, the oscillation amplitudes of $\rho_\mathrm{max}(t)/\rho_\mathrm{max}(0)$ and $\mathcal{B}_\mathrm{max}(t)/\mathcal{B}_\mathrm{max}(0)$ are reduced by a factor of $e^{-1}$ at $t\sim 8.5$ ms and $t\sim 6$ ms respectively.

We illustrate the radial profiles of the rest-mass density $\rho(r)$ (top panel) and the toroidal magnetic field strength $\mathcal{B}_{\phi}(r)$ (bottom panel) in the equatorial plane (i.e. polar angle $\theta= \pi/2$) at 
$t=0$ ms (gray solid lines), 0.5 ms (red dash-dotted lines), 1.0 ms (orange dash-dotted lines), 5.0 ms (yellow dash-dotted lines), 10.0 ms (green dash-dotted lines), and 20.0 ms (blue dotted lines) in Fig. \ref{fig2}.
These profiles show that the \emph{phase-transition-induced collapse} causes both both $\rho(r)$ and $\mathcal{B}_{\phi}(r)$ of the star initially oscillates around a new equilibrium configuration, and the resulting hybrid star eventually obtains a slightly higher central rest-mass density and maximum magnetic field strength. 
In addition, the magnetic field inside the star becomes more concentrated towards the core with a shift of the maximum magnetic field strength position $r_{\mathcal{B}}$ (dashed lines) to smaller values.
Furthermore, new configurations of $\rho(r)$ and $\mathcal{B}_{\phi}(r)$ are obtained at $t=10$ ms and remain until at least $t=20$ ms.
This agrees with the fact that the oscillation amplitudes of $\rho_\mathrm{max}(t)/\rho_\mathrm{max}(0)$ and $\mathcal{B}_\mathrm{max}(t)/\mathcal{B}_\mathrm{max}(0)$ have already decreased by a factor of $e^{-1}$ at $t\sim 8.5$ ms and $t\sim 6$ ms respectively as discussed in Fig. \ref{fig1}.

\begin{figure*}
    \centering
    \includegraphics[width=\textwidth, angle=0]{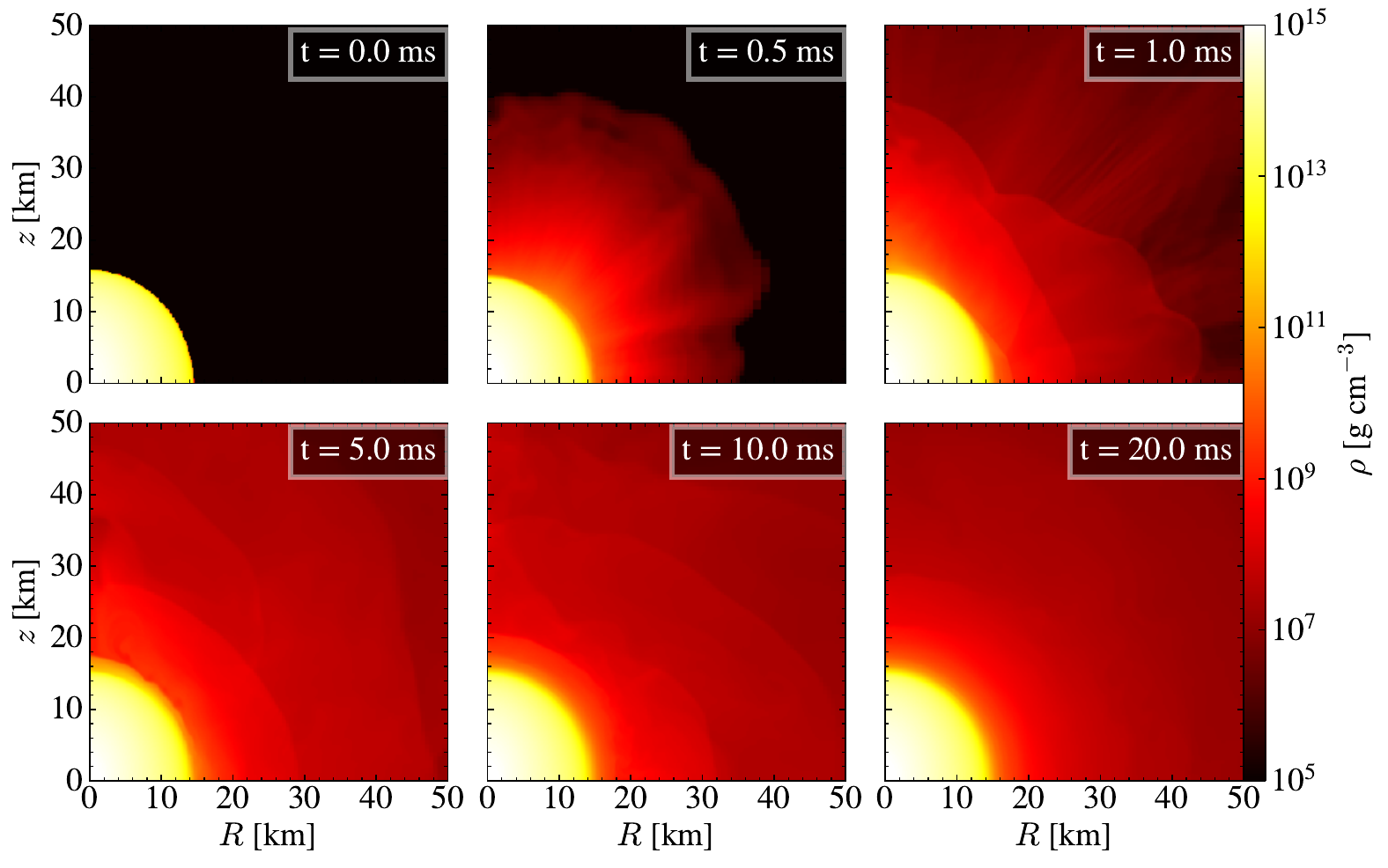}
    \caption{
	        The 2D profile of the rest-mass density $\rho$ for the simulation with the initial model T1K6 and the exponent $\delta=3$ at $t=0$ ms (upper left), 0.5 ms (upper middle), 1.0 ms (upper right), 5.0 ms (lower left), 10.0 ms (lower middle), and 20.0 ms (lower right).
            Model T1K6 has an initial maximum magnetic field strength $\mathcal{B}_\mathrm{max} = 5.52 \times 10^{17}$ G and $\delta$ is an exponent describing the pressure contribution due to quark matter in the mixed phase.
            We observe the ejection of matter from the star into its surrounding 'atmosphere' during its evolution. 
            By defining the material within the atmosphere as possessing a rest-mass density $\rho$ equal to or less than $10^{-2}$ of the lower threshold density $\rho_\mathrm{hm}$ (i.e. $\rho \leq 6.97 \times 10^{12}$ g cm$^{-3}$), we find that only around 0.1\% of the initial baryonic mass $M_{0}(0)$ of the star is expelled into the atmosphere after 20 ms of evolution.
            Therefore, the overall shape of the 2D profile of $\rho$ within the star preserves well throughout the evolution.
	         }
    \label{fig3}	
\end{figure*}

\begin{figure*}
    \centering
    \includegraphics[width=\textwidth, angle=0]{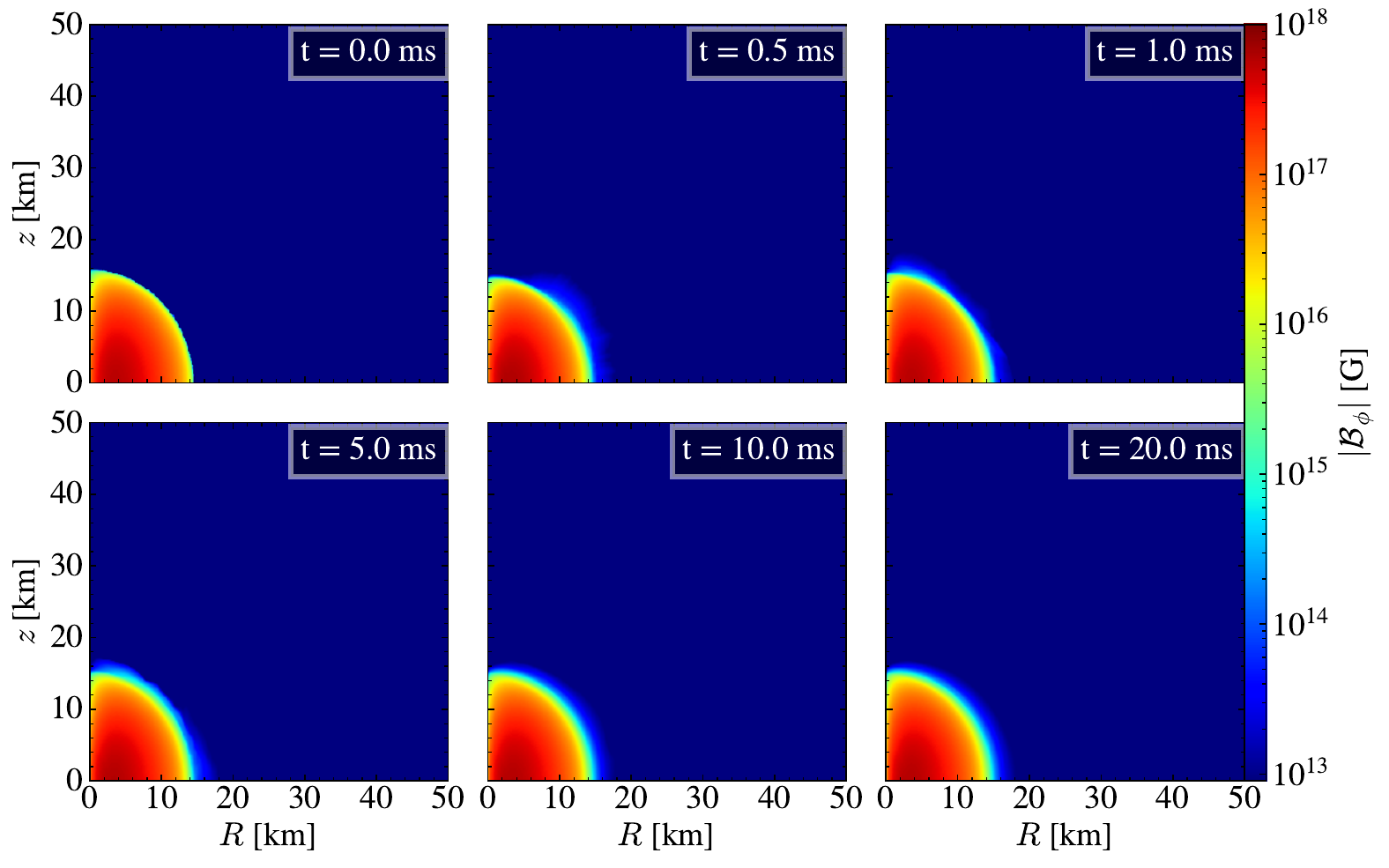}
    \caption{
	        The 2D profile of the absolute value of the toroidal magnetic field strength $|\mathcal{B}_{\phi}|$ for the simulation with the initial model T1K6 and the exponent $\delta=3$ at $t=0$ ms (upper left), 0.5 ms (upper middle), 1.0 ms (upper right), 5.0 ms (lower left), 10.0 ms (lower middle), and 20.0 ms (lower right).
            Model T1K6 has an initial maximum magnetic field strength $\mathcal{B}_\mathrm{max} = 5.52 \times 10^{17}$ G and $\delta$ is an exponent describing the pressure contribution due to quark matter in the mixed phase.
            We observe that the overall shape of the 2D profile of $|\mathcal{B}_\phi|$ within the star preserves well throughout the evolution.
	         }
    \label{fig4}	
\end{figure*}

We also plot the 2D profiles of the rest-mass density $\rho$ in Fig. \ref{fig3} and the absolute value of the toroidal magnetic field strength $|\mathcal{B}_\phi|$  in Fig. \ref{fig4} at $t=0$ ms (upper left), 0.5 ms (upper middle), 1.0 ms (upper right), 5.0 ms (lower left), 10.0 ms (lower middle), and 20.0 ms (lower right).
In particular, the 2D profile of $\rho$ in Fig. \ref{fig3} shows that some matter is ejected from the star into its surrounding `atmosphere' during its evolution.
To estimate the amount of ejected matter, we define the matter within the atmosphere as having a rest-mass density $\rho$ equal to or less than $10^{-2}$ of the lower threshold density $\rho_\mathrm{hm}$ (i.e. $\rho \leq 6.97 \times 10^{12}$ g cm$^{-3}$). 
We find that only about 0.1\% of the initial baryonic mass $M_{0}(0)$ of the star is expelled into the atmosphere after 20 ms of evolution.
Accordingly, we observe that the overall shapes of the 2D profiles of both $\rho$ and $|\mathcal{B}_\phi|$ within the star preserve well throughout the evolution.

\section{Properties of the resulting magnetized hybrid stars}\label{sec:properties}
As discussed in Sections \ref{sec:intro} and \ref{sec:num_method}, although the magnetic effects on polytropic stars (e.g. \citealt{2014MNRAS.439.3541P}) and hybrid stars (e.g. \citealt{2016MNRAS.456.2937F}) have already been studied, the stellar properties are only studied through equilibrium modeling.
In addition, it is not apparent how a \emph{phase-transition-induced collapses} affects the configurations of a magnetized star (e.g. profiles of rest-mass density and magnetic field).
Thus, it is necessary to investigate whether the magnetic effects found by equilibrium modeling are still valid for describing the magnetized hybrid stars that are dynamically formed by \emph{phase-transition-induced collapses}.

\begin{figure*}
    \centering
    \includegraphics[width=\textwidth, angle=0]{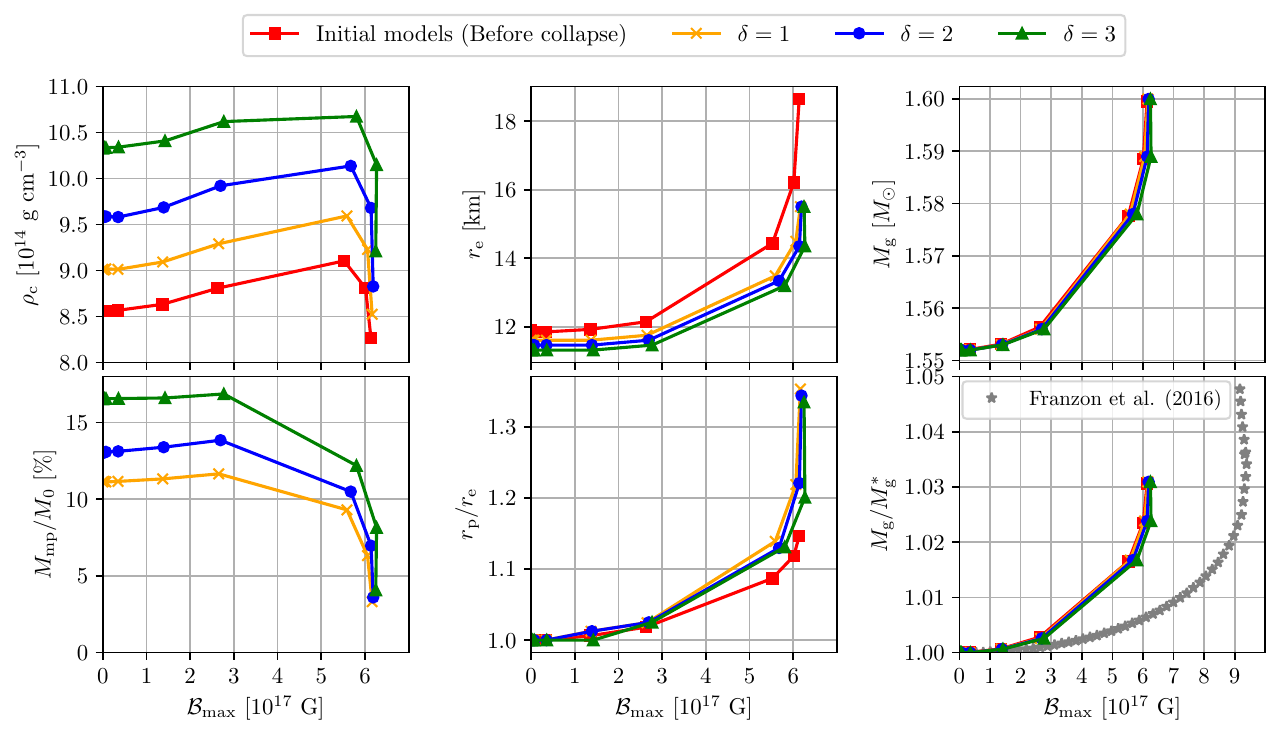}
    \caption{
            Plots of various stellar quantities against the maximum magnetic field strength $\mathcal{B}_\mathrm{max}$ in our hybrid star models: central rest-mass density $\rho_\mathrm{c}$ (top left panel), baryonic mass fraction in the mixed phase $M_\mathrm{mp} / M_{0}$ (bottom left panel), equatorial radius $r_\mathrm{e}$ (top middle panel), polar-to-equatorial radius ratio $r_\mathrm{p} / r_\mathrm{e}$ (bottom middle panel), and gravitational mass $M_\mathrm{g}$ (top right panel).
            Data points are shown for 3 sequences with $\delta \in \{1, 2, 3\}$ and compared with initial pre-collapse neutron star models (red squares), where $\delta$ signifies the pressure contribution from quark matter in the mixed phase.
            Quantities in magnetized hybrid star models vary similarly with $\mathcal{B}_\mathrm{max}$ as initial models, showing insensitivity for $\mathcal{B}_\mathrm{max} \lesssim 3 \times 10^{17}$ G and notable changes for $\mathcal{B}_\mathrm{max} \gtrsim 5 \times 10^{17}$ G.
            Thus, the magnetic effects on the stellar properties of the initial neutron star equilibrium models still accurately describe the resulting magnetized hybrid stars after the \emph{phase-transition-induced collapses}.
            In addition, $\delta$ minimally affects $r_\mathrm{e}$, $r_\mathrm{p}/r_\mathrm{e}$, and $M_{g}$ across all $\mathcal{B}_\mathrm{max}$ values. In contrast, $\rho_\mathrm{c}$ and $M_\mathrm{mp} / M_{0}$ noticeably increase with $\delta$ for $\mathcal{B}_\mathrm{max} \lesssim 3 \times 10^{17}$ G, becoming less sensitive for $\mathcal{B}_\mathrm{max} \gtrsim 5 \times 10^{17}$ G.
            Furthermore, we compare the normalized gravitational mass $M_\mathrm{g}/M_\mathrm{g}^{*}$ against $\mathcal{B}_\mathrm{max}$ (bottom right panel) with models from \citet{2016MNRAS.456.2937F} (gray stars), where $M_\mathrm{g}^{*}$ represents the gravitational mass of non-magnetized reference models.
            $M_\mathrm{g}/M_\mathrm{g}^{*}$ increases similarly with $\mathcal{B}_\mathrm{max}$ in our models and those of \citet{2016MNRAS.456.2937F}.
            Hence, we find agreement across different methods in supporting the magnetic effects on hybrid stars.}
    \label{fig5}	
\end{figure*}

Hence, to better examine the magnetic effects on the properties of our resulting magnetized hybrid star models, we plot in Fig. \ref{fig5} different microscopic and macroscopic quantities against the maximum magnetic field strength $\mathcal{B}_\mathrm{max}$ of the stars. 
The data points of our resulting magnetized hybrid star models formed from \emph{phase-transition-induced collapses} are arranged into 3 sequences with 3 values of $\delta \in \{1,2,3\}$, where $\delta$ is an exponent quantifying the pressure contribution due to quark matter in the mixed phase. 
Here, we define the equatorial radius and polar radius of the resulting hybrid stars as the radial positions where the rest-mass density $\rho$ is less than or equal to $10^{-2}$ of the lower threshold density $\rho_\mathrm{hm}$ (i.e. $\rho \leq 6.97 \times 10^{12}$ g cm$^{-3}$).
We also plot the data points of our initial neutron star models before the phase transition collapses (red squares) as a comparison.

We find that all microscopic and macroscopic quantities of the resulting magnetized hybrid star models vary with the $\mathcal{B}_\mathrm{max}$ in similar ways as those of the initial neutron star models before the \emph{phase-transition-induced collapses}.
Specifically, for $\mathcal{B}_\mathrm{max} \gtrsim 5 \times 10^{17}$ G, all microscopic and macroscopic quantities vary strongly with $\mathcal{B}_\mathrm{max}$, irrespective of $\delta$.
When $\mathcal{B}_\mathrm{max} \lesssim 3 \times 10^{17}$ G, all quantities vary slightly with $\mathcal{B}_\mathrm{max}$.
Hence, these behaviors confirm that the magnetic effects on the initial neutron star equilibrium models still accurately describe their corresponding magnetized hybrid star models formed from \emph{phase-transition-induced collapses}.

Here, we highlight how the magnetic field affects different microscopic and macroscopic quantities of a star in detail for completeness.
On the one hand, the central rest-mass density $\rho_\mathrm{c}$ (top left panel) and the baryonic mass fraction of the matter in the mixed phase $M_\mathrm{mp} / M_{0}$ (bottom left panel) decrease with $\mathcal{B}_\mathrm{max}$. 
These decreasing behaviors could be understood in terms of magnetic pressure. 
As the magnetic pressure becomes more dominant due to the increasing $\mathcal{B}_\mathrm{max}$, matter is pushed off-center to a greater extent. 
As a result, the rest-mass density $\rho$ in the stellar core reduces, giving a smaller $\rho_\mathrm{c}$. 
Moreover, as described in Eq. (\ref{eqn6}), reducing $\rho$ in the core contributes to a smaller fraction of matter that undergoes the phase transition and thus gives a smaller $M_\mathrm{mp} / M_{0}$.

On the other hand, the equatorial radius $r_\mathrm{e}$ (top middle panel), the polar radius to equatorial radius ratio $r_\mathrm{p} / r_\mathrm{e}$ (bottom middle panel) and the gravitational mass $M_\mathrm{g}$ (top right panel) all increase with $\mathcal{B}_\mathrm{max}$. 
The increase in $M_\mathrm{g}$ is due to the increasing contribution of the magnetic field to $M_\mathrm{g}$ (corresponds to the increasing $\mathcal{B}^2$ term of Eq. (B1) in \citealt{2014MNRAS.439.3541P} for example). 
The other two increasing trends could be interpreted in terms of magnetic deformation. As the matter is pushed off-center by the increasing magnetic pressure, both $r_\mathrm{p}$ and $r_\mathrm{e}$ increase and the star then deviates from spherical symmetry. 
Previous studies of magnetized neutron star equilibrium models (e.g. \citealt{2008PhRvD..78d4045K,2009ApJ...698..541K,2012MNRAS.427.3406F}) indicate that a purely toroidal field deforms the stars to prolate shape, corresponding to $r_\mathrm{p} / r_\mathrm{e} > 1$. 
Thus, increasing $\mathcal{B}_\mathrm{max}$ of the toroidal field in our models causes the increase in $r_\mathrm{p} / r_\mathrm{e}$. 

We also examine the effect of pressure contribution due to quark matter $\delta$ on different quantities of the hybrid stars.
$\delta$ has a negligible effect on $r_\mathrm{e}$, $r_\mathrm{p}/r_\mathrm{e}$ and $M_{g}$ for all values of $\mathcal{B}_\mathrm{max}$. 
On the contrary, $\rho_\mathrm{c}$ and $M_\mathrm{mp} / M_{0}$ increase substantially with $\delta$ for $\mathcal{B}_\mathrm{max} \lesssim 3 \times 10^{17}$ G but they become less sensitive to $\delta$ for $\mathcal{B}_\mathrm{max} \gtrsim 5 \times 10^{17}$ G. 
These increasing trends could be interpreted in relation to pressure reduction. 
With the increasing value of $\delta$, the contribution due to quark matter to the total pressure becomes more important and the pressure reduction is enlarged. 
As a result, this enlarged pressure reduction makes the star collapse to a configuration with a higher $\rho_\mathrm{c}$ and $M_\mathrm{mp} / M_{0}$.

We compare our resulting hybrid stars with \citet{2016MNRAS.456.2937F}. 
The magnetized hybrid star models considered in this study are also in axisymmetry, but the magnetic field is purely poloidal. 
A poloidal field would make the stars oblate instead of prolate. 
Also, these models have a different baryonic mass $M_{\mathrm{0}}=2.2$ $M_{\odot}$. 
These equilibrium models are constructed by solving the coupled Maxwell–Einstein equations. 
They also employed a more realistic equation of state with both magnetic and thermal effects taken into account.

We plot the normalized gravitational mass $M_\mathrm{g}/M_\mathrm{g}^{*}$ against $\mathcal{B}_\mathrm{max}$ (bottom right panel) to compare with the models computed in \citet{2016MNRAS.456.2937F} (gray stars), where $M_\mathrm{g}^{*}$ is the gravitational mass of the non-magnetized reference models.
We observe that $M_\mathrm{g} / M_\mathrm{g}^{*}$ increases with $\mathcal{B}_\mathrm{max}$ similarly for the models in our simulations and \citet{2016MNRAS.456.2937F}.
Besides, this previous study also found that the magnetic field reduces the central baryon number density and hinders the appearance of matter in quark and mixed phases. 
These also agree with the trends of $\rho_\mathrm{c}$ and $M_\mathrm{mp} / M_{0}$ for our models.
Accordingly, despite the disparity in field geometry, baryonic mass, and construction method, our models agree qualitatively with the models in the previous study.
This similarity provides additional support that the properties of hybrid the properties of the magnetised hybrid stars presented here are robust.

\section{Conclusions}\label{sec:conclusions}
In this paper, we studied the formation of a magnetized hybrid star by performing 2D axisymmetric general-relativistic magnetohydrodynamics simulations. 
We first found that the maximum values of rest-mass density and magnetic field strength in the stars rise slightly after a phase transition. 
The magnetic field also becomes more concentrated towards the center. In addition, the magnetic field and the rest-mass density are coupled during the process. 
We then investigated the properties of the resulting magnetized hybrid stars. 
Both macroscopic and microscopic quantities of the hybrid stars are not sensitive to the magnetic field until $\mathcal{B}_\mathrm{max} \gtrsim 5 \times 10^{17}$ G, where all quantities change significantly. 
Specifically, the magnetic deformation decreases the rest-mass density dramatically, leading to a substantial reduction in the matter fraction in the mixed phase. 
Similar trends for these quantities are found compared with \citet{2016MNRAS.456.2937F}.

This work takes the first step to dynamically studying magnetized hybrid stars. 
Several natural extensions should be considered to model them more realistically. 
First, a more realistic equation of state, which includes thermal and magnetic effects, should be adopted. 
In addition, since magnetized stars with purely toroidal fields are expected to be unstable, the suppression of instability in this work is mainly due to the restriction to 2D axisymmetry. 
The effects of purely poloidal fields and the twisted torus configurations should also be investigated. 
Since these field geometries extend to the outer region of neutron stars, a force-free/resistive magnetohydrodynamics solver is necessary for more realistic modeling. 
Also, 3D simulations without axisymmetry should be conducted to include the instability of magnetic fields. 
Finally, as most observations suggested that neutron stars rotate, rotations should also be included in future studies.

\section*{Acknowledgements}

We acknowledge the support of the CUHK Central High-Performance Computing Cluster, on which the simulations in this work have been performed. 
This work was partially supported by grants from the Research Grants Council of Hong Kong (Project No. CUHK 14306419), the Croucher Innovation Award from the Croucher Foundation Hong Kong, and the Direct Grant for Research from the Research Committee of The Chinese University of Hong Kong. 
P.C.-K.C. acknowledges support from NSF Grant PHY-2020275 (Network for Neutrinos, Nuclear Astrophysics, and Symmetries (N3AS)).

\section*{Data Availability}

The data underlying this article are available in the article.



\bibliographystyle{mnras}
\bibliography{references} 




\appendix

\section{Resolution study}\label{sec:res_study}
\begin{figure}
    \centering
    \includegraphics[width=\columnwidth, angle=0]{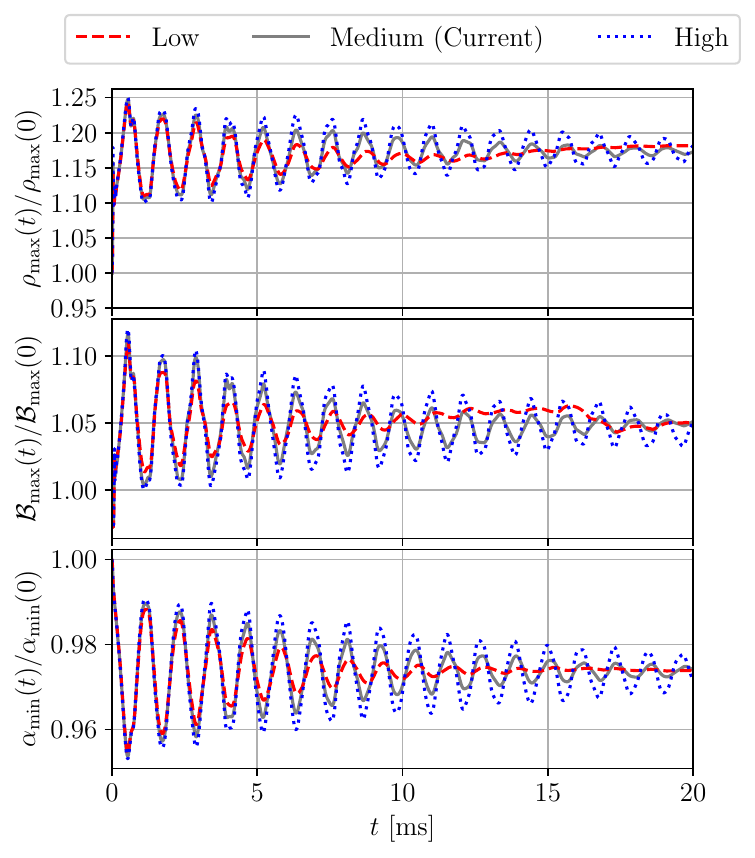}
    \caption{
            The time evolution of the maximum rest-mass density $\rho_\mathrm{max}(t)$ (top panel), the maximum magnetic field strength $\mathcal{B}_\mathrm{max}(t)$ (middle panel) and the minimum lapse function $\alpha_\mathrm{min}(t)$ (bottom panel) relative to their initial values $\rho_\mathrm{max}(0)$, $\mathcal{B}_\mathrm{max}(0)$ and $\alpha_\mathrm{min}(0)$ for the simulation with the initial model T1K6 and the exponent $\delta=3$ at low (red dashed line), medium (gray solid line), and high (blue dotted line) resolutions.
            The medium resolution here is the resolution used in the simulations of the current study.
            We observe that all three quantities at different refinement levels undergo damped oscillation similarly, and they all eventually obtain a new equilibrium value, which is slightly higher than the initial value.
            The difference in the refinement level only affects the damping timescales and slightly changes the final equilibrium values of these two quantities.
            In particular, a lower refinement level corresponds to a higher numerical dissipation and thus gives a shorter damping timescale.
            Therefore, the numerical dissipation does not affect the qualitative behavior of the star in our simulations.            
    }
    \label{figa1}	
\end{figure}

Apart from the shock heating effect mentioned in Section \ref{sec:dynamics}, numerical dissipation also plays a role in the damping of the oscillation of the star in our simulations.
Here, we demonstrate that numerical dissipation does not affect the simulation results qualitatively.
Fig. \ref{figa1} shows the time evolution of the maximum rest-mass density $\rho_\mathrm{max}(t)$ (top panel), the maximum magnetic field strength $\mathcal{B}_\mathrm{max}(t)$ (middle panel) and the minimum lapse function $\alpha_\mathrm{min}(t)$ (bottom panel), relative to their initial values, $\rho_\mathrm{max}(0)$, $\mathcal{B}_\mathrm{max}(0)$ and $\alpha_\mathrm{min}(0)$ for the simulation with the initial model T1K6 and the exponent $\delta=3$ at low (red dashed line), medium (gray solid line), and high (blue dotted line) resolutions.
The medium resolution here is the resolution used in the simulations of the current study.
Since AMR is used in our simulations, the grid spacings for both $R$ and $z$ ($\Delta R$ or $\Delta z$) are not uniform but are assigned according to the AMR level of each region.
The region containing the star has the finest grid spacing ($\Delta R_\mathrm{min}$ and $\Delta z_\mathrm{min}$) of the whole domain.
The low, medium, and high resolutions allow 5, 6, and 7 AMR levels in the domain, respectively. 
These resolutions also correspond to the finest grid spacing $\Delta R_\mathrm{min} = \Delta z_\mathrm{min} \approx$ 290 m, 140 m, and 70 m, respectively.
We observe that all three quantities at different refinement levels undergo damped oscillation similarly, and they all eventually obtain a new equilibrium value, which is slightly higher than the initial value.
The difference in the refinement level only affects the damping timescales and slightly changes the final equilibrium values of these two quantities.
In particular, a lower refinement level corresponds to a higher numerical dissipation and thus gives a shorter damping timescale.
Therefore, the numerical dissipation does not affect the qualitative behavior of the star in our simulations.

\section{Finite time for the initial phase transition}\label{sec:finite_pt}
\begin{figure}
    \centering
    \includegraphics[width=\columnwidth, angle=0]{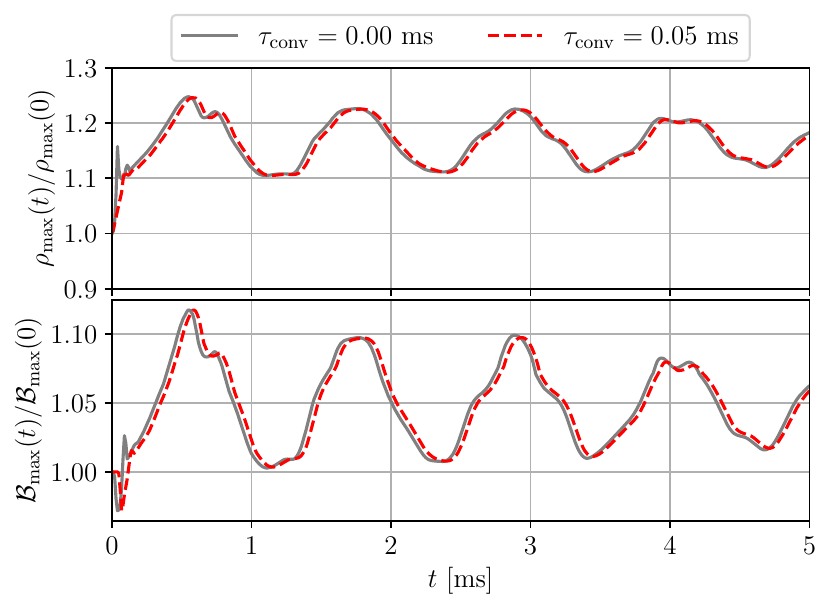}
    \caption{The first 5 ms time evolution of the maximum values of the rest-mass density $\rho_\mathrm{max}(t)$ (top panel) and the magnetic field strength $\mathcal{B}_\mathrm{max}(t)$ (bottom panel) relative to their initial values $\rho_\mathrm{max}(0)$ and $\mathcal{B}_\mathrm{max}(0)$, for the simulation with the initial model T1K6 and the exponent $\delta=3$ for two cases: one with an instantaneous initial phase transition (i.e. $\tau_\mathrm{conv}=0.00$ ms) (gray solid line) and the other one with $\tau_\mathrm{conv}=0.05$ ms (red dashed line). 
    The current study assumes an instantaneous initial phase transition, and $\tau_\mathrm{conv}=0.05$ ms corresponds to a phase-transition timescale in the order of a detonation-like conversion.
    We observe that the finite $\tau_\mathrm{conv}$ does not change the qualitative behavior of both quantities and only produces a slight time lag.
    Therefore, the qualitative dynamics of phase-transition-induced collapses remain unaffected by imposing a finite timescale in the order of a detonation-like conversion for the initial phase transition even in the presence of a magnetic field.
    }
    \label{figb1}	
\end{figure}

In \citet{2009MNRAS.392...52A}, it has been shown that imposing a finite timescale in the order of a detonation-like conversion for the initial phase transition (i.e. $\tau_\mathrm{conv}=0.05$ ms) does not affect the dynamics of phase-transition-induced collapses qualitatively and only contributes a slight time lag for the case without a magnetic field.
Here, we show that imposing such a finite timescale for the initial phase transition does not affect the simulation results qualitatively even with a magnetic field.
Fig. \ref{figb1} illustrates the first 5 ms time evolution of the maximum values of the rest-mass density $\rho_\mathrm{max}(t)$ (top panel) and the magnetic field strength $\mathcal{B}_\mathrm{max}(t)$ (bottom panel) relative to their initial values $\rho_\mathrm{max}(0)$ and $\mathcal{B}_\mathrm{max}(0)$.
We consider a simulation with the initial model T1K6 and an exponent $\delta=3$, comparing two cases: one with an instantaneous initial phase transition (i.e. $\tau_\mathrm{conv}=0.00$ ms) (gray solid line) and the other with a finite timescale of $\tau_\mathrm{conv}=0.05$ ms (red dashed line).
The current study assumes an instantaneous initial phase transition and the value of $\tau_\mathrm{conv}=0.05$ ms corresponds to a timescale of phase transition in the order of a detonation-like conversion.
We observe that the finite $\tau_\mathrm{conv}$ does not change the behavior of both quantities qualitatively and only produces a slight time lag.
Consequently, imposing a finite timescale for the initial phase transition in the order of a detonation-like conversion does not qualitatively affect the dynamics of phase-transition-induced collapses even when there is a magnetic field.

\bsp	
\label{lastpage}
\end{document}